**Exotic Spintronic Properties of Transition-metal Monolayers on Graphyne**

*Xiaoxiong Ren, Junsheng Huang, Ping Li, Yun Zhang, and Zhi-Xin Guo*\*

Xiaoxiong Ren, Junsheng Huang, Dr. Ping Li, Prof. Zhi-Xin Guo
State Key Laboratory for Mechanical Behavior of Materials, Center for Spintronics and Quantum System, School of Materials Science and Engineering, Xi'an Jiaotong University, Xi'an, Shaanxi, 710049, China.

Prof. Yun Zhang
Department of Physics and Information Technology, Baoji University of Arts and Sciences, Baoji 721016, China.

Prof. Zhi-Xin Guo
Key Laboratory of Polar Materials and Devices, Ministry of Education.

E-mail: zxguo08@xjtu.edu.cn

Abstract text: The recent discovery of two-dimensional (2D) magnetic materials which are compound of transition metal (TM) with other elements, has opened new avenues for basic research on low-dimensional magnetism and potential applications in spintronics. To further explore new 2D magnets of pure TM is an interesting topic. Based on the first-principles calculations, here we propose a strategy of obtaining monolayer TM magnets, i.e., depositing TM atoms on graphyne (Gy) which has proper hexagonal hollow geometry. We find that TM monolayer with perfect hexagonal geometry can be formed on Gy. The TM monolayer exhibits a wealth of physical properties in dependence of TM species, such as ferromagnetic and antiferromagnetic ground states, as well as intriguing semimetal and half-metal characteristics. We also find that the half-metal characteristics makes the monolayer TM have great potential applications in the horizontal magnetic tunnel junction (MTJ) devices, where the tunneling magnetoresistance can reach as high as 850000%. Our results provide a new framework for obtaining 2D magnets with outstanding spintronic properties.

## 1. Introduction

Two-dimensional (2D) magnetic materials can exhibit many novel physical properties which are particular useful in the spintronics.[1] In the past decades, magnetic 2D materials were usually created in an extrinsic way, e.g. defect engineering [2] on non-magnetic 2D materials. Until recently, the discovery of 2D van der Waals (vdW) materials with intrinsic magnetism opens a new avenue to the 2D magnets. The interplay of dimensionality, correlation, charge, orbital character, and topology makes 2D magnetic crystals and heterostructures extremely fertile condensed matter systems with a large reservoir of exotic properties, such as quantum anomalous hall effect,[3] unique spin-orbit coupling effect,[4] strong tunneling magnetoresistance (TMR) effect[1d, 5] and so on. These fascinating properties make them have great potential applications in the future spintronics.

At present, most studies focus on 2D magnetic materials composed of multiple compounds of transition metal (TM) and other elements, e.g. $CrI_3$,[6] $MnBi_2Te_4$[7] and $Fe_3GeTe_2$[8] and so on. Despite the fascinating properties of these 2D magnetic materials, the appearance of multiple element components easily induces defects such as uneven distribution of components and structures[9] during the synthesis process, which severely limits their wide application in spin-polarized devices. Therefore, it is greatly disable to find a way to obtain 2D magnetic material composed of pure TM.

On the other hand, the TM magnetic films such as Co, Fe are mostly used magnetic materials in the present spintronic devices.[10] In order to meet the development needs of miniaturization and high performance of spintronic devices, many methods had been proposed to further reduce the thickness of TM magnetic film[11] as well as to synthesize them on a semiconductor substrate.[12] However, owing to the agglomeration nature of TM atoms and the complex surface structure of semiconductor, it is extremely hard to obtain an atomic-thick magnetic TM film on the semiconductor substrates.

## 2. Results and Discussion:

Here we propose that the 2D magnetic material can be obtained by depositing TM atoms on Gy which has a hexagonal hollow geometry. The deposited TM atoms prefer to locate on the hollow site of Gy and form a perfect honeycomb monolayer. We find several TM elements such as V that are non-magnetic (NM) in their bulk phases have AFM ground state in their 2D structure, while the usually believed FM materials such as Ni presents NM ground state. The variation of magnetism can be attributed to the combined effects of out-of-plane symmetry broken and orbital hybridization with Gy. We also find that Nb, Mo exhibit the 2D half-metal characteristic, where the tunneling magnetoresistance in the magnetic tunnel junction (MTJ) is estimated as high as 850000%.

We have considered γ-Gy as the substrate, which consists of hexagonal carbon rings and acetylene linkages, with 6 C atoms forming the $C_{sp} \equiv C_{sp}$ hybridization and the remaining 6 C atoms forming the $C_{sp^2}-C_{sp}$ hybridization, respectively.[13] Our DFT calculations show that the lattice constant of freestanding Gy is 6.890 Å, agree with previous studies.[14] We first systematically explored the stable geometries of the entire TMs from 3d to 5d on Gy. A common feature is that all the TM atoms prefer to locate on the hollow site above the center of acetylenic ring, where the monolayer TM with graphene-like structure can be formed in a proper coverage (**Figure 1**a). Considering that the nonbonded interaction generally appears between atoms with distances of more than 2 times larger of their bonded length, and the nearest-neighbor distances between two TM atoms are around 3.8 Å, only about 30% larger than the bond lengths in their bulk phases, it is natural to expect that there is still strong bonding interaction between them. As discussed below, this feature can lead to strong Heisenberg exchange interaction and thus the FM phase of TM monolayers. On the other hand, the strong interface interaction between the TM and C induces significant buckling of graphyne (d1 in **Table 1**), the value of which depends on the specific TM element (highly related to the atomic radius).

**Table I.** Lattice constants and structure information of selected FM and AFM systems. a represents the lattice constants of the systems. M represents the magnetic moment of a single TM in the Gy/TM systems. The d1 and d2 represent the distance between the transition metal atomic layer and the first layer C and the distance between the two layers C, respectively. $E_c$ represents the cohesive energy of TM monolayer on Gy. S means semiconductor, M means metal, and HM means half-metal.

| System | a (Å) | M ($\mu_B$/TM) | d1 (Å) | d2 (Å) | Bandgap (eV) | $E_c$ (meV) | Physical properties |
|---|---|---|---|---|---|---|---|
| Gy/V | 6.566 | 1.096 | 0.77 | 0.82 | 0.345 | 3.352 | S |
| Gy/Nb | 6.589 | 0.606 | 0.98 | 0.83 | 0.066 | 4.931 | S |
| Gy/Ta | 6.588 | 0.642 | 0.93 | 0.86 | 0.367 | 5.940 | S |
| Gy/Cr | 7.035 | 3.366 | 0.00 | 0.00 | —— | 2.117 | M |
| Gy/Mo | 6.585 | 1.582 | 0.92 | 0.81 | —— | 3.512 | HM |
| Gy/W | 6.552 | 1.456 | 0.89 | 0.85 | —— | 4.898 | HM |
| Gy/Mn | 6.752 | 3.018 | 0.58 | 0.61 | 0.852 | 2.389 | S |
| Gy/Tc | 6.632 | 1.721 | 0.84 | 0.75 | 0.398 | 4.240 | M |
| Gy/Re | 6.586 | 1.521 | 0.84 | 0.80 | 0.135 | 3.992 | M |
| Gy/Fe | 6.905 | 2.380 | 0.40 | 0.30 | —— | 3.379 | M |
| Gy/Co | 6.946 | 1.293 | 0.12 | 0.12 | —— | 4.013 | M |

To evaluate the stability, we calculated the cohesive energy $E_c$ of monolayer TM on Gy which is defined as,[15]

$$E_c = (E_{Gy} + N_{TM}\mu_{TM} - E_{tot})/N_{TM}$$

where $E_{Gy}$ and $E_{tot}$ are the total energies of Gy and Gy/TM, respectively. $N_{TM}$ is the number of TM atoms on graphyne and $\mu_{TM}$ is the chemical potential of TM which is adopted as the energy of an isolated TM atom. The cohesive energy $E_c$ is defined above is the energy gain to grow TM monolayer on the Gy surface. As shown in Table I and **Table SI**, $E_c$ of all TM monolayers is above 2.0 eV except for five group I B and II B elements (Ag, Au, Zn, Cd, Hg) which have completely filled orbitals. Such $E_c$ value is larger than that of a monolayer TM on metal substrate and that of a group-IV monolayer on TM substrate, both of which had been experimentally synthesized, [16] showing the possibility of experimental synthesis for the monolayer TM. Moreover, in 2014 Zhao et al[17] had successfully trapped the Fe atoms in the pores of graphene, which finally form the single-atom-thick Fe membranes. These 2D Fe nanomembranes are directly imaged and are shown to have a square lattice at room temperature. The system in our work is similar to graphene/Fe, the experimental growth of transition metal (TM) monolayer on Gy is expected to be also realized in the near future. The dynamical stability of such system was also confirmed in our previous study for Gy/Hf via ab initio molecular dynamics (AIMD) simulations. [18]

Figure 1b shows the ground states of TM monolayers on Gy in the form of an element periodic table. As one can see, TM monolayers present plentiful stable phases covering FM, AFM and NM. Note that all the FM and AFM phases are for TM elements of group III B-VIII B with partially filled *d* orbitals. We further define weak-FM (AFM) and strong-FM (AFM) phases for a FM (AFM) monolayer with magnetic moment smaller and larger than 0.5 $\mu_B$/atom, respectively. In addition, we have evaluated the effect of Hubbard U on the DFT results. Considering that the value of Hubbard U depends on both the structure of the system and the types of TM atoms, and there is not available experimental result for us to fit a reliable value of U in Gy/TM system, we simply adopted U = 2.0 eV for all the Gy/TM systems with strong FM and AFM phases. As shown in Table SII, the Hubbard U does not significantly change the magnetic moment as well as the magnetic ground state in our studied system.

As shown in Figure 1b, all the elements of group V B-VII B have a strong FM/AFM phase, whereas the ones of group III B, IV B and a part of group VIII B present the weak-FM/AFM phase. It is noticed that only four of VIII B elements have FM or AFM phase, while the remaining five present the NM phase.

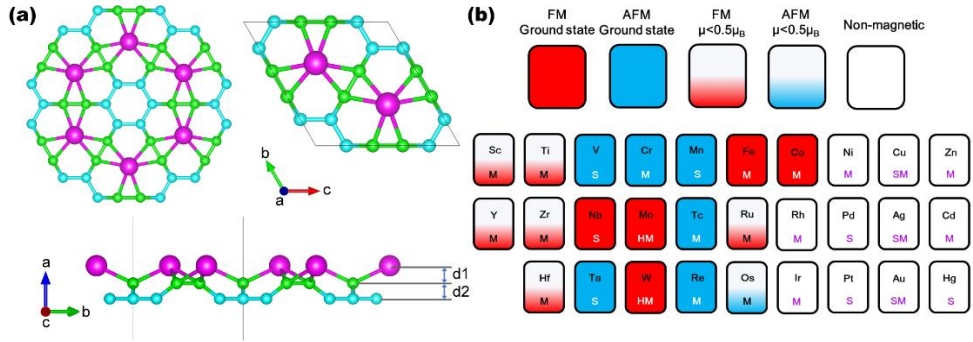

**Figure 1.** (Color online) Calculated structure and physical characteristics of Gy/TM system. (a) Hexagonal structure formed by TM in Gy/TM system (upper left), as well as the top view (upper right) and side view (lower) of a unit cell Gy/TM structure. The magenta, green, and blue atoms represent the TM atom, C atom in the nearest neighbor, and C atom in the next nearest neighbor to TM atom, respectively. d1 and d2 represent the distance between the TM atomic layer and the first layer C and the distance between the two layers C, respectively. (b) The physical properties of Gy/TM are given in the form of the periodic table, where the red, blue, and white background represent the magnetic ground states of ferromagnetic (FM), antiferromagnetic (AFM), and non-magnetic (NM), respectively. The full and partial filling of the background color represent that the magnetic moment of a single TM is greater than and less than 0.5 $\mu_B$, respectively. The letters S, M, HM and SM represent the properties of semiconductors, metals, half-metals and semimetals respectively.

In order to understand the origin of magnetism in TM monolayers, we carried out the d-orbital projection density of states (PDOS) analysis on TM atoms (**Figure 2**a-c). It is found that the spilt of energy levels of *d* orbitals (except for Fe and Mn as discussed below) basically present a triangular-prism-crystal-field (TPCF)-like distribution, i.e., the degenerate $d_{x^2-y^2}/d_{xy}$ orbitals, $d_{z^2}$ orbital and degenerate $d_{xz}/d_{yz}$ orbitals hold the energy levels from low to high energy, respectively. The formation of such TPCF-like distribution is a result of strong hybridization between *d* orbitals of TM atoms and *p* orbitals of the underlying C atoms as indicated in **Figure S1**, which significantly lowers the orbital symmetry of TM atoms compared to their bulk phases. This feature is confirmed by the calculated shortest TM–C distances (about 2.2 Å), which is comparable to the sum of the covalent atomic radii of TM and C atoms. Moreover, we find that the hybridization with *p* orbitals can help to fill the empty *d* orbitals by about two electrons per TM atom. Therefore, the variation of magnetism in the 2D structure can be owing to the strong *p-d* orbital hybridizations that induce significant orbital-symmetry breaking and orbital filling, which are responsible for the distribution of spin-up and spin-down electrons in the crystal field. Considering that both the atomic radii and number of *d* electrons affecting *p-d* orbital hybridizations vary with the type of TM element, it is reasonable to appear of plentiful stable phases for the TM monolayers on Gy, depending not only on the group index but also on the period index. Note that, all the TM elements with six *d*-orbital electrons, i.e., Ni, Pd, Pt, present the NM phase in the Gy/TM structure, because the hybridization with *p* orbitals of C makes *s* and *d* orbitals completely filled.

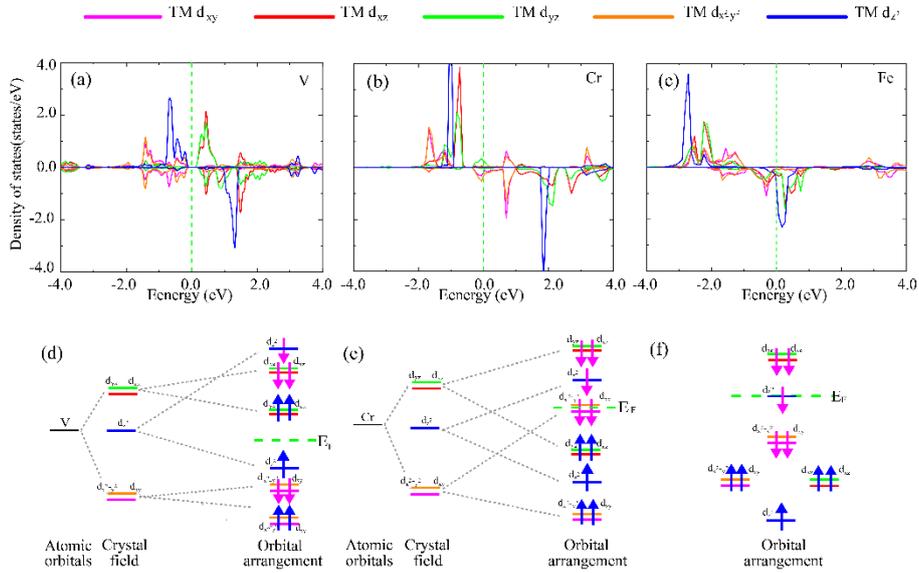

**Figure 2**. (Color online) The calculated PDOS and *d* orbital arrangement of a single TM atom in the Gy/TM system. The *d*-orbital PDOS (a) and orbital arrangement (d) of a V atom in Gy/V with larger atomic radius (≥1.34Å). The orbital arrangement characteristics for Gy/TM with TM=Cr, Mn, Te, Ta, Re are similar to Gy/V. (b)(c) and (e)(f) are the results for systems with smaller atomic radius (<1.30 Å). Note that Gy/Co has similar orbital arrangement with Gy/Cr, and Gy/Mn has similar orbital arrangement with Gy/Fe. The red and green lines, the blue lines, the magenta and orange lines represent degenerate $d_{x^2-y^2}/d_{xy}$ orbitals, single $d_{z^2}$ orbitals, degenerate $d_{xz}/d_{yz}$ orbitals, respectively. $E_F$ is Fermi level. The spilt of energy levels of *d* orbitals (except for Fe and Mn as discussed in the text) basically present a TPCF-like distribution.

Then we focus on the 11 TM monolayers which have the strong FM or AFM ground states with larger magnetic momentum (> 0.5 $\mu_B$/atom). As shown in Figure 1b, 5 of 11 (Fe, Co, Nb, Mo, W) TM elements present the strong FM ground state, and the remaining 6 (V, Cr, Mn, Te, Ta, Re) have the strong AFM ground state. The PDOS calculations show that these ground states can be described by three kinds of orbital arrangement (Figure 2d-f). The first two are originated from the TPCF-like filed, that is, the TM elements with atomic radius larger than 1.34 Å (V, Nb, Ta, Mo, W, Tc, Re) have $d_{z^2}$ and $d_{xz}/d_{yz}$ orbitals locating on the VBM and CBM respectively, whereas those with atomic radius smaller than 1.30 Å (Cr, Co) mainly have $d_{x^2-y^2}/d_{xy}$ orbitals nearby $E_F$. The third kind is for Fe and Mn which also have small atomic radius (1.26 Å and 1.27 Å). While, the exchange field presents complex orbital arrangement with degenerated $d_{x^2-y^2}/d_{xy}$ and $d_{xz}/d_{yz}$ orbitals as shown in Figure 2f, which cannot be directly derived from the TPCF-like filed. These results show that the exchange field of TM monolayer is induced by the combined effect of atomic radii (more dominating) and number of $d$ valance electrons of the TM element.

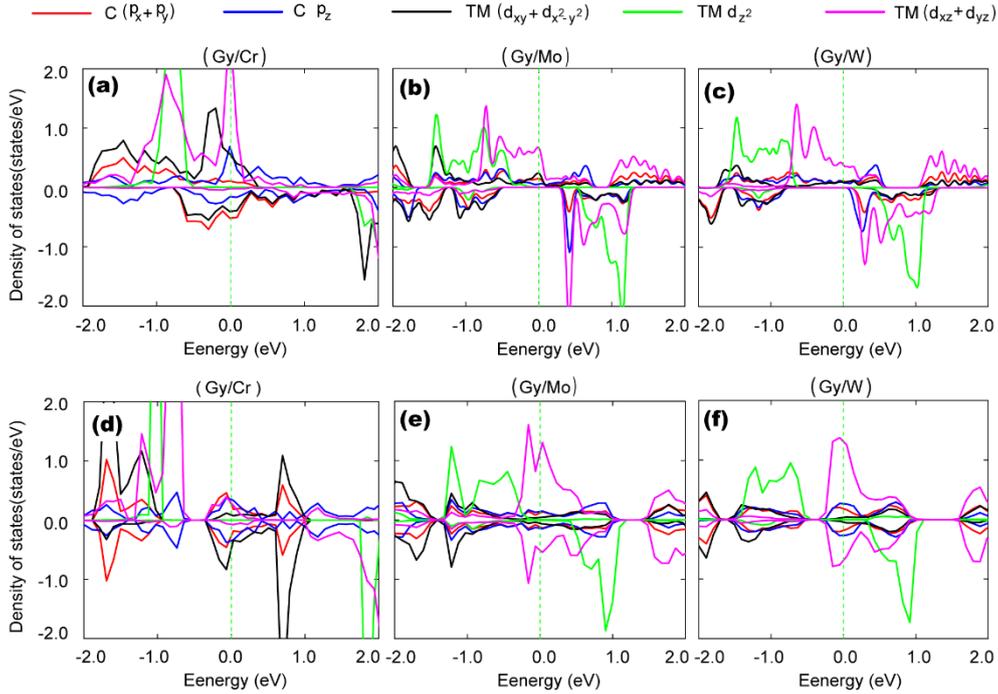

**Figure 3**. (Color online) The calculated PDOS of Gy/TM. (a), (b) and (c) represent the PDOS of magnetic systems of Gy/TM(TM=Cr, Mo, W). (d), (e) and (f) represent the PDOS of antiferromagnetic systems of them, respectively. The red (for degenerate $p_x/p_x$ orbital) and blue (for $p_z$ orbital) lines represent the PDOS of the six C nearest to TM, respectively. Black lines represent the degenerate $d_{x^2-y^2}/d_{xy}$ orbitals, magenta lines represent the degenerate $d_{xz}/d_{yz}$ orbitals, and green lines represent the $d_{z^2}$ orbitals of TM, respectively. The competition between the super-exchange induced by the hybridization of $p$-$d$ orbitals and the direct exchange induced by the overlaps between the *d-d* orbitals determines the magnetic ground state of the systems. The PDOS of *d* orbitals are from one TM atom of the unit cell.

Based on the detailed analysis of PDOS, we found that the ground FM/AFM magnetic state is attributed to the competition between direct exchange and super-exchange of *d* electrons. In the following, we show the details by taking Cr, Mo, W in group VI B as an example. We firstly considered Cr monolayer, which has an AFM ground state. **Figure 3**a and 3d show the PDOS of a Cr atom in the FM and AFM phases, respectively. It is shown that there are obvious overlaps of PDOS peaks for $p_x/p_y$ and $d_{x^2-y^2}/d_{xy}$, as well as for $p_z$ and $d_{xz}/d_{yz}$ orbitals, indicating the significant *p-d* orbital hybridizations. Such orbital hybridizations are expected to induce strong super-exchange magnetic interactions contributing to the AFM phase, between two nearest Cr atoms bridged by the two bonded C atoms (Figure 1). On the other hand, there are also direct *d-d* orbital overlaps between two nearest-neighbor Cr atoms, contributing to the FM phase. Therefore, the magnetic ground state is determined by the competition between the super-exchange and direct exchange effects, where the super-exchange effect is dominant for Cr monolayer. Note that the direct exchange effect would be dominant for Mo and W monolayers, because the *p-d* orbital hybridizations contributing to the AFM phase become weaker, whereas the direct *d-d* orbital overlaps become stronger due to the increase of atomic radii. This feature is confirmed by the calculated PDOS of a Mo (W) atom shown in Figure 3b and 3e (Figure 3c and 3f) with both FM and AFM phases, where the overlapped peaks of *p-d* orbitals become obviously smaller but the width of *d* orbital DOS nearby Fermi level gets much larger compared to that of Cr. We have additionally calculated and analyzed the spin chare density of Gy/Cr, Gy/Mo and Gy/W ground states (Figure S5), and found that the results are consistent with the PDOS analysis.

In addition, we have calculated the exchange constant J for the Gy/TM with the magnetic moment larger than 0.5 μ B per TM atom (Table SII). It is found that the exchange constant of some structures, i.e., Gy/Mo and Gy/W, reaches around -30 meV, meaning that their Curie temperatures are at room-temperature level. To insure this, we have further explored the Curie temperatures of Gy/Mo and Gy/W. The calculation was performed by using the Monte Carlo method, where the corresponding parameters are obtained by the DFT calculations. The calculation results are shown in Figure S4. It is found that the Curie temperatures of Gy/Mo and Gy/W are 243K and 293K, respectively, which are pretty high in comparation with the usual 2D magnetic materials[6,7].

Now we come to discuss the electronic properties of Gy/TM. As shown in Figure 1b and **Figure S3**, the TM monolayers exhibit plentiful electronic properties, including metal, semiconductor, as well as semimetal. Particularly, the FM and AFM systems are of great interest for potential device applications, and here we focus on the TM monolayers with FM ground state. **Figure 4** shows the calculated band structures of Nb, Mo, and W monolayers on Gy, which exhibit either half-metal or half-semiconductor characteristics distinguished from their bulk phases. As shown in the Figure 4, Mo and W monolayers are ideal half-metals with pure spin-up electrons appearing in [-0.5, 0.4] eV and [-0.8, 0.1] eV, respectively. Whereas, Nb monolayer exhibits half-semiconductor characteristic with energy gap of 0.1 eV, where the pure spin-up electrons appears in [-0.5, 0.2] eV. On the other hand, although the Fe and Co monolayers do not have half-metal characteristic, they still have pure spin-down electrons nearby the Fermi level, i.e., [-1.0, -0.2] eV and [-0.7, -0.3] eV, respectively (**Figure S2**). This feature shows these TM monolayers can be ideal materials for the spin transport devices.

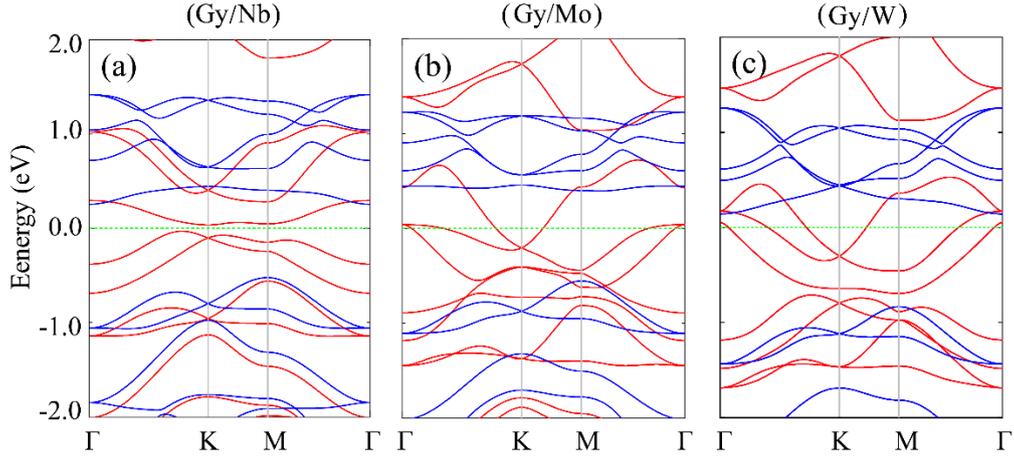

**Figure 4.** (Color online) Half-metallic Gy/TM system. (a), (b), and (c) respectively represent the half-metal system of Gy/TM(TM=Nb, Mo, W). Red and blue represent the spin-up and spin-down electronic states, respectively. Gy/Nb exhibits half-semiconductor characteristic with energy gap of 0.1 eV, where the pure spin-up electrons appears in [-0.5, 0.2] eV. Mo and W systems are half-metals with pure spin-up electrons appearing in [-0.5, 0.4] eV and [-0.8, 0.1] eV, respectively.

To verify the application potentials of TM monolayers in the spintronic devices, we further explored the TMR of a horizontal MTJ device based on the Gy/TM structure (**Figure 5**a). It is known that the vertical MTJ devices based on 2D materials have been widely studied in the past decade, however, the investigation on horizontal MTJ is still scarce due to the difficulty in realizing the 2D horizontal magnetic heterostructures in experiments. Here we propose that one can realize the horizontal MTJ device much easier in the use of the Gy/TM system, i.e., deposit different TM atoms using the MASK method to make magnetic layers, and leave underposited region (bare Gy) as the tunneling layer. Note that, the bare Gy is a semiconductor with band gap about 0.4 eV, which makes the undeposited Gy region ideal candidate for the tunneling layer.[19]

We further show a horizontal MTJ device model in Figure 5a, where the source and drain electrodes are composed of Gy/Mo, and the free magnetic layer and tunneling layer are composed by Gy/W and bare Gy, respectively. The parallel (P) and anti-parallel (AP) magnetization states of the MTJ are realized by flipping the magnetization direction of W monolayer as indicated in Figure 5b. The calculated transmission spectrum of the P and AP states for different spins with zero bias are shown in Figure 5c. As one can see, in both cases the spin-up ($T_\uparrow$) electron transmission is larger than that of spin-down ($T_\downarrow$) electrons nearby the Fermi level by about ten orders, manifesting a perfect spin-filtering effect. We further evaluated the TMR of the MTJ, which is defined as | ($T_P$-$T_{AP}$)/$T_{AP}$ |×100%, with $T_P$= $T_\uparrow$ + $T_\downarrow$ and $T_{AP}$= $T_\uparrow$ + $T_\downarrow$, respectively. As shown in Figure 5d, the 2D device has exceptionally TMR of about 2800% at the Fermi level and 850000% at -0.63eV, respectively. Especially, the 850000% is significantly larger than the reported record realized in the vertical MTJ based on multilayer $CrI_3$ (57000%).[20]

The unusually high TMR can be attributed to the half-metal property of Gy/W and Gy/Mo, which induces much larger transmission probability of spin up electrons than that of the spin down electrons in Gy/W, with the Gy/Mo region being in the spin up state (Figures 5b and 5c). Due to the spin transfer torque effect, the transmission probability of majority spins in the P state is several orders of magnitude higher than at of AP state, which results in the high TMR for the MTJ. We have also explored the I-V curve (Figure S7). It is found that the current density of them under 0.1 V are obviously larger than that in the vertical MTJ based of multilayer $CrI_3$ [20]. This result indicates that the horizontal Gy/TM type MTJ is more superior in the low-power spintronic devices.

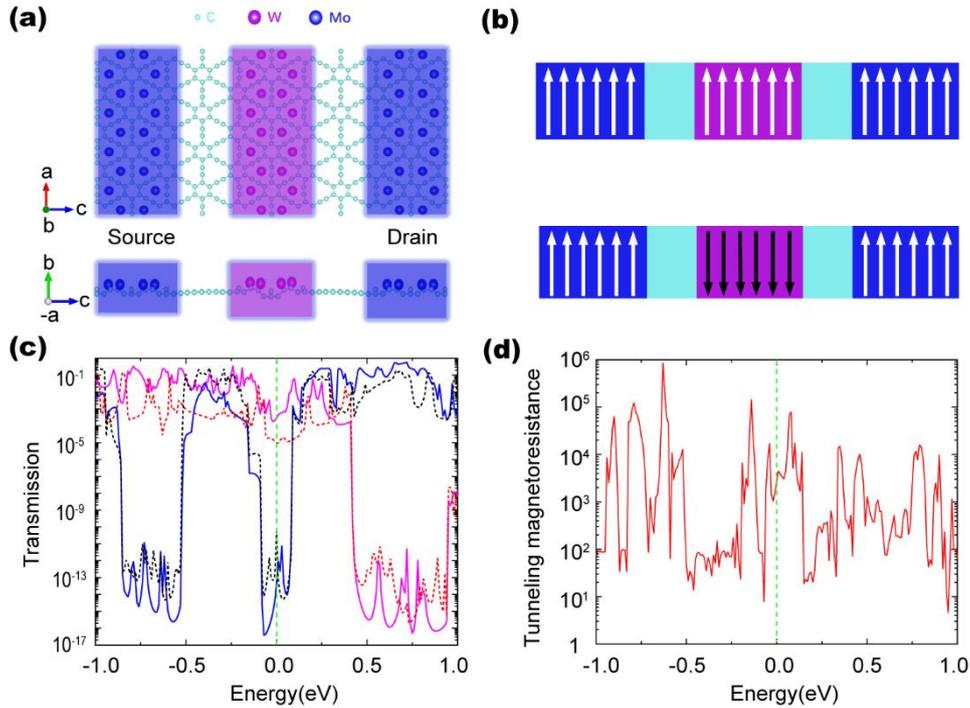

**Figure 5**. (Color online) Gy/TM horizontal MTJ device model and calculated electronic transport properties. (a) MTJ device model in top view (upper panel) and side view (lower panel). (b) Schematic diagram of the magnetic sequence of the device changing from a parallel (P) state to an anti-parallel (AP) state. (c) Calculated transmission spectra of the P state and AP state under zero bias. The solid line and the dashed line represent the device configuration in the P and AP states, respectively. The magenta (red) and blue (black) solid lines represent the transmission for spin up and spin down states, respectively. (d) Calculated tunneling magnetoresistance as a function of energy level.

## 3. Conclusion

In summary, by using the first-principles calculations we have proposed a new approach to obtain monolayer TM magnets, i.e., depositing TM atoms on substrate with hexagonal hollow geometry. We have demonstrated the feasibility of this approach on Gy substrate, where the TM monolayer with perfect hexagonal geometry has been formed. We have also found the TM monolayer exhibits a wealth of physical properties, including FM and AFM ground states, as well as semimetal and half-metal characteristics. The competition between direct exchange and super-exchange interactions have been proposed to explain the variation of magnetic phases. Finally, we have demonstrated the TM monolayers have great potential applications in the spintronics, where the TMR in the horizontal MTJ device model is calculated to be 2800% at the Fermi level and as high as 850000% at -0.63eV, respectively.

## 4. Methods

The first-principles calculations were performed by using the density functional theory (DFT)-based Vienna ab initio simulation package (VASP).[21] The ion–electron interaction was treated by the projector augmented-wave (PAW) technique.[22] Exchange–correlation energies were taken into account by the generalized gradient approximation (GGA) using the Perdew–Burke–Ernzerhof functional.[23] For TM monolayer, periodic boundary condition for the 2D structure was used, and a vacuum layer of 15 Å was set to avoid interactions in the out-of-plane direction. A plane wave basis set with a cutoff energy of 450 eV was used, and the first Brillouin-zone integration was carried out by a 15×15×1 Γ-centered Monkhorst-Pack grid.[24] Both of the atomic positions and lattice constants were optimized using a conjugate gradient method with criteria of energy and Hellmann–Feynman force convergence being less than $10^{-6}$ eV per unit cell and 0.01 eV Å$^{-1}$, respectively.

As for the calculation of tunneling magnetoresistance in the MTJ, we first optimized the atomic positions using a conjugate gradient method with criteria of energy and Hellmann–Feynman force convergence being less than $10^{-5}$ eV per unit cell and 0.02 eV Å$^{-1}$, respectively. Then we constructed the MTJ device model as shown in Figure 4a. The transmission calculations were carried out using the Atomistic Simulation Toolkit (ATK) with the PBE pseudopotentials distributed in the QuantumWise package.[25] The transmission was calculated using the non-equilibrium Green's function (NEGF) approach.[26] A Gy/W/Gy supercell was used as the scattering region, ideally attached on both sides to semi-infinite Gy/Mo leads.[27] The Dirichlet boundary condition in the c-direction and periodic boundary conditions in the a-b plane were used in the simulation (Figure 4a) and the length of electrodes are 6.712Å. The temperature was set at 300K. The convergence tested (Figure S6) Monkhorst-Pack grid ($15 \times 1 \times 100$) was used with a mesh cutoff energy (55 Hartree) for the electrodes and central region.

The Medium basis set was adopted with PseudoDojo pseudopotentials. The NEGF–DFT self-consistency was controlled by a numerical tolerance of $10^{-5}$ eV.

**Supporting Information**
Supporting Information is available from the Wiley Online Library or from the author.

**Acknowledgements**

We are grateful for useful discussions with Dr. Lei Wang and Dr. Yongliang Shi. This work is supported by the National Natural Science Foundation of China (No. 12074301 and No. 12004295), National Key R&D Program of China (2018YFB0407600), Science Fund for Distinguished Young Scholars of Hunan Province (No. 2018JJ1022), Fundamental Research Funds for Central Universities (No. xzy012019062), and Open Research Fund of Key Laboratory of Polar Materials and Devices, Ministry of Education. P.L. thanks China's Postdoctoral Science Foundation funded project (NO. 2020M673364).

Received: ()
Revised: ()
Published online: ()

# Supporting Information

**Exotic Spintronic Properties of Transition-metal Monolayers on Graphyne**

*Xiaoxiong Ren, Junsheng Huang, Ping Li, Yun Zhang, and Zhi-Xin Guo*[*]

**Contents:**
Table 1 The structure and physical properties of systems Gy/TM
Table II The Magnetic properties of Gy/TM systems.
S1 Band projection of Gy/TM system
S2 The band structures of the Gy/Fe and Gy/Co systems
S3 Band structure of Gy/TM in different ground states
S4 The curie temperature of Gy/Mo and Gy/W
S5 Spin charge density and super exchange interaction diagrams of Gy/Cr, Gy/Mo and Gy/W systems.
S6 K-point test of Gy/Mo and Gy/W device
S7 I-V curve of the Gy/Mo and Gy/W device.

**Table 1. The structure and physical properties of systems Gy/TM**

**Table S1 Lattice constants and structure information of the other research systems.** a represents the lattice constants of the systems. m represents the magnetic moment of a single TM in the Gy/TM systems. The d1 and d2 represent the distance between the transition metal atomic layer and the first layer C and the distance between the two layers C, respectively. Ec represents the cohesive energy of TM monolayer on Gy.

| System | a (Å) | m (μB/TM) | d1 (Å) | d2 (Å) | Bandgap (eV) | Ec (eV) |
|---|---|---|---|---|---|---|
| GY/Sc | 6.780 | 0.091 | 1.08 | 0.60 | —— | 3.509 |
| GY/Ti | 6.595 | 0.105 | 0.87 | 0.81 | —— | 4.247 |
| GY/Ni | 6.928 | —— | 0.03 | 0.00 | —— | 4.054 |
| GY/Cu | 6.958 | —— | 0.12 | 0.06 | —— | 2.284 |
| GY/Zn | 6.874 | —— | 3.70 | 0.00 | —— | 0.066 |
| GY/Y  | 6.791 | 0.199 | 1.40 | 0.58 | —— | 3.469 |
| GY/Zr | 6.579 | 0.069 | 1.16 | 0.83 | —— | 4.929 |
| GY/Ru | 6.698 | 0.409 | 0.76 | 0.66 | —— | 4.322 |
| GY/Rh | 6.847 | —— | 0.66 | 0.45 | —— | 4.200 |
| GY/Pd | 7.034 | —— | 0.26 | 0.17 | 0.451 | 2.674 |
| GY/Ag | 6.890 | —— | 1.72 | 0.12 | —— | 0.616 |
| GY/Cd | 6.869 | —— | 3.78 | 0.00 | —— | 0.097 |
| GY/Hf | 6.553 | 0.120 | 1.11 | 0.86 | —— | 4.890 |
| GY/Os | 6.604 | 0.222 | 0.80 | 0.76 | 0.005 | 4.759 |
| GY/Ir | 6.760 | —— | 0.69 | 0.60 | —— | 4.674 |
| GY/Pt | 6.985 | —— | 0.38 | 0.34 | 0.695 | 3.976 |
| GY/Au | 7.048 | —— | 0.50 | 0.23 | —— | 0.641 |
| GY/Hg | 6.874 | —— | 3.90 | 0.00 | 0.448 | 0.053 |

**Table SII The magnetic properties of Gy/TM systems.** M1 indicates that the magnetic moment of a single magnetic atom in the Gy/TM systems of Hubbard U is not considered, and M2 is the case where Hubbard U is 2.0eV. Considering whether the Hubbard U has no effect on the magnetic ground state of the systems. The energies of the ferromagnetic (FM), antiferromagnetic (AFM) and non-magnetic (NM) states of the Gy/TM system are shown in the table. J represents the exchange constant between two magnetic atoms, and it is expressed as the energy of the ferromagnetic state minus the energy of the antiferromagnetic state.

| System | M1 (no U) (μB/TM) | M2 (U=2.0eV) (μB/TM) | Magnetic ground state (no U) | Magnetic ground state (U=2.0eV) | Energy (NM) (eV) | Energy (FM) (eV) | Energy (AFM) (eV) | J (meV) |
|---|---|---|---|---|---|---|---|---|
| GY/V  | 1.096 | 1.151 | AFM | AFM | -116.1399 | -116.8348 | -116.8370 | 0.366 |
| GY/Nb | 0.606 | 0.664 | FM  | FM  | -118.8891 | -119.2591 | -119.2465 | -2.106 |
| GY/Ta | 0.642 | 0.786 | AFM | AFM | -121.4746 | -121.8006 | -121.8603 | 9.949 |
| GY/Cr | 3.366 | 3.841 | AFM | AFM | -115.4337 | -117.9823 | -118.1581 | 0.652 |
| GY/Mo | 1.582 | 1.790 | FM  | FM  | -118.4707 | -119.2546 | -118.9615 | -30.062 |
| GY/W  | 1.456 | 1.772 | FM  | FM  | -121.1446 | -121.9228 | -121.6550 | -27.468 |
| GY/Mn | 3.018 | 3.351 | AFM | AFM | -115.0803 | -118.0071 | -118.0429 | 0.546 |
| GY/Tc | 1.721 | 2.256 | AFM | AFM | -117.4897 | -117.8874 | -118.0847 | 10.352 |
| GY/Re | 1.521 | 1.714 | AFM | AFM | -119.7986 | -120.1762 | -120.2549 | 4.066 |
| GY/Fe | 2.380 | 2.696 | FM  | FM  | -114.3529 | -115.9935 | -115.9664 | -0.725 |
| GY/Co | 1.293 | 1.298 | FM  | FM  | -113.6174 | -114.1272 | -113.8015 | -24.127 |

**S1. Band projection of Gy/TM system**

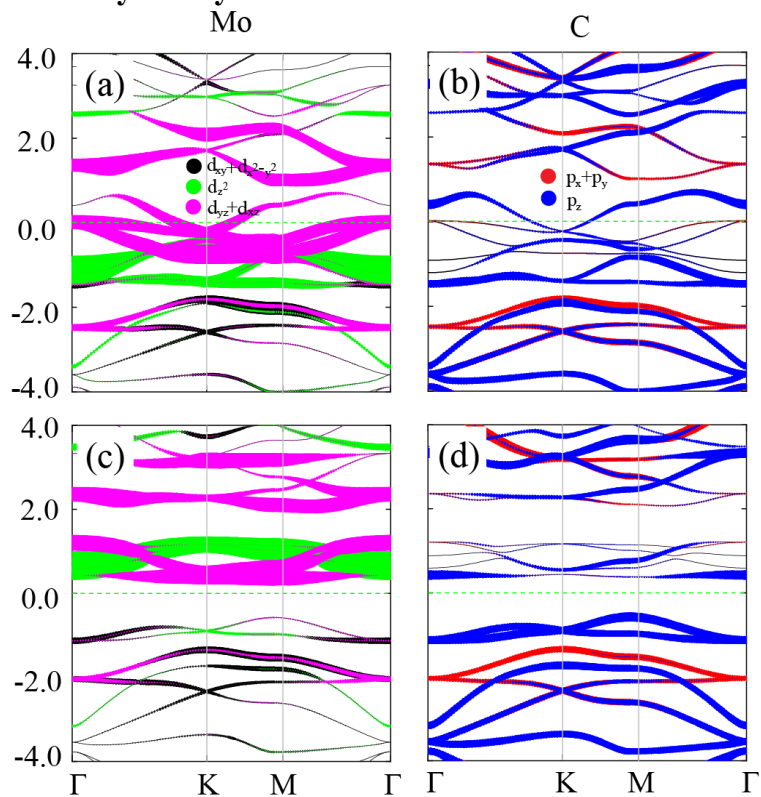

**Figure S1. Projection band structure of Gy/Mo.** Calculated band structures of Gy/Mo projected onto Mo (a and c) and C (b and d) with different orbital symmetries. Figures (a) and (b) are for spin-up electrons of Mo and C, whereas, Figures (c) and (d) are for spin-down electrons of Mo and C, respectively. Note that, the orbital hybridization features presented for Gy/Mo is also applicable to other TM elements in the Gy/TM system.

**S2. The band structures of the Gy/Fe and Gy/Co systems**

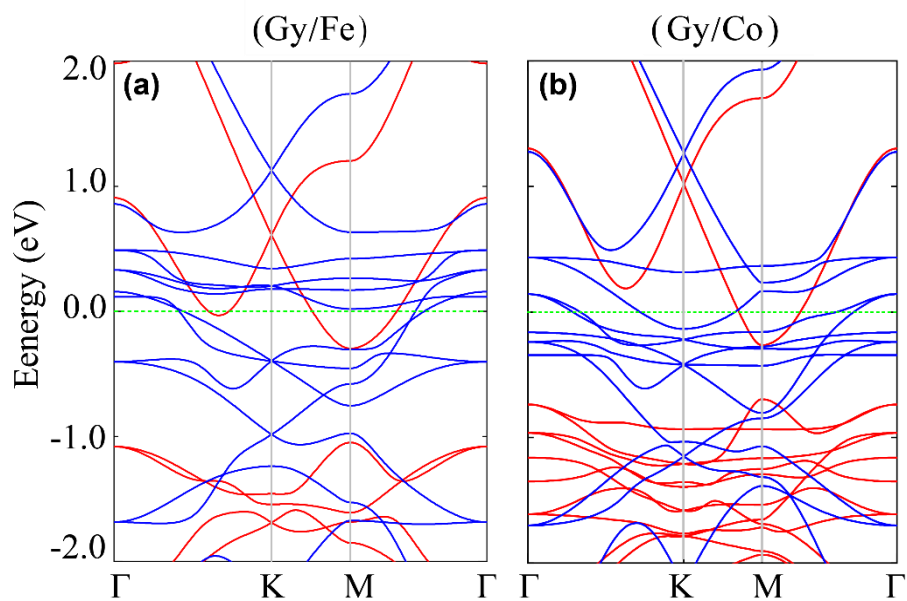

**Figure S2. The band structure of Gy/Fe and Gy/Co system.** (a) and (b) represent the system of Gy/Fe and Gy/Co, respectively. Red and blue represent the spin-up and spin-down electronic states, respectively. They have pure sin-down electrons nearby the Fermi level, i.e., [-1.0, -0.2] eV and [-0.7, -0.3] eV, respectively

## S3. Band structure of Gy/TM in different ground states

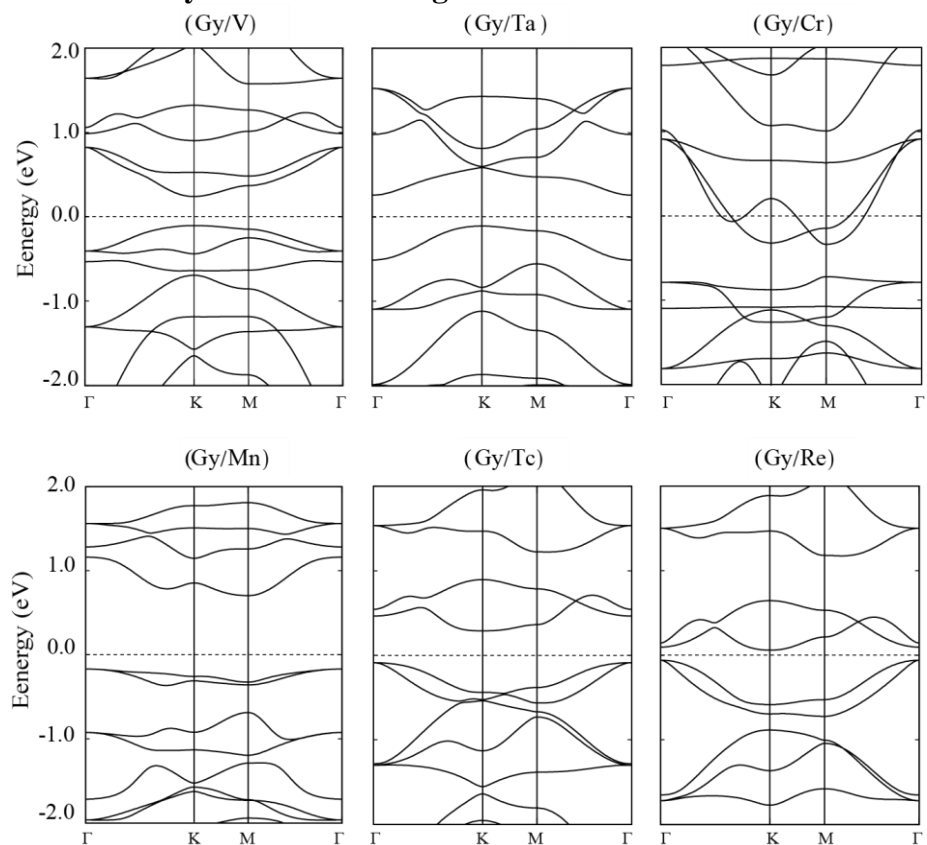

**Figure S3.1 The band structure of the antiferromagnetic ground state with magnetic moment of a single magnetic atom is more than 0.5μB.** It can be seen that, except Gy/Cr is metal, the rest are semiconductors

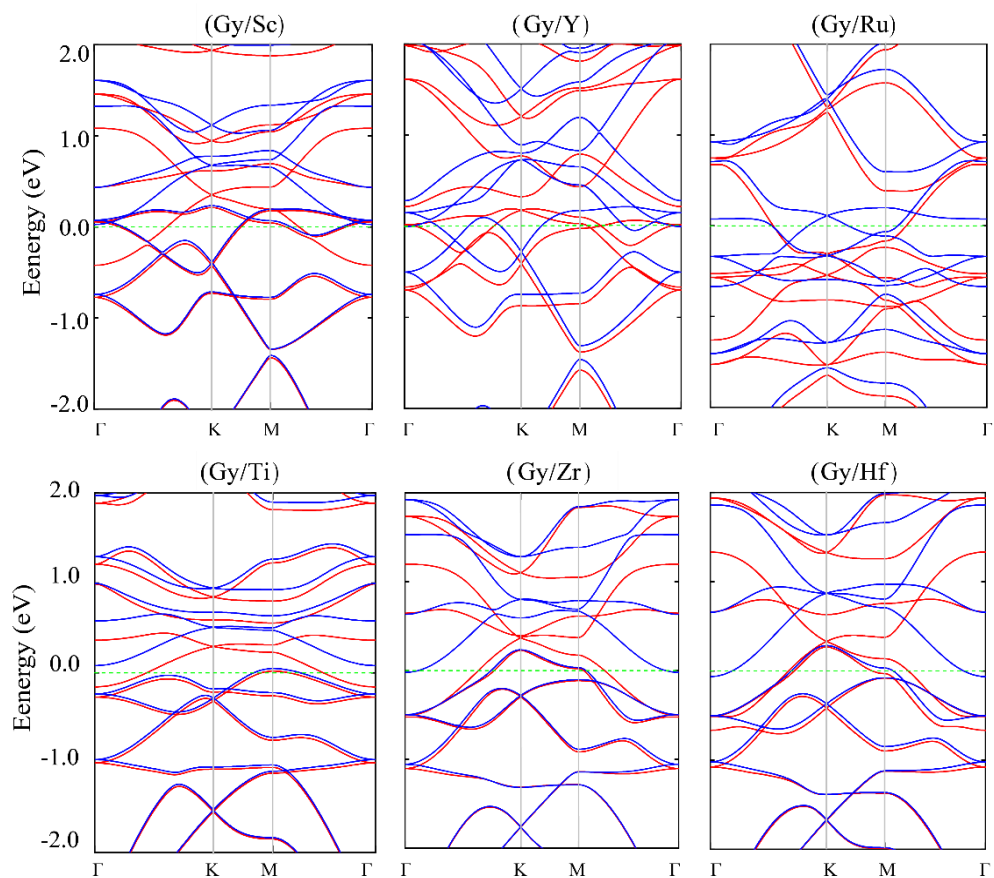

**Figure S3.2 The band structure of the ferromagnetic ground state with magnetic moment of a single magnetic atom is less than 0.5µB.** They are all ferromagnetic ground state. Red and blue lines represent the spin-up and spin-down electronic states, respectively.

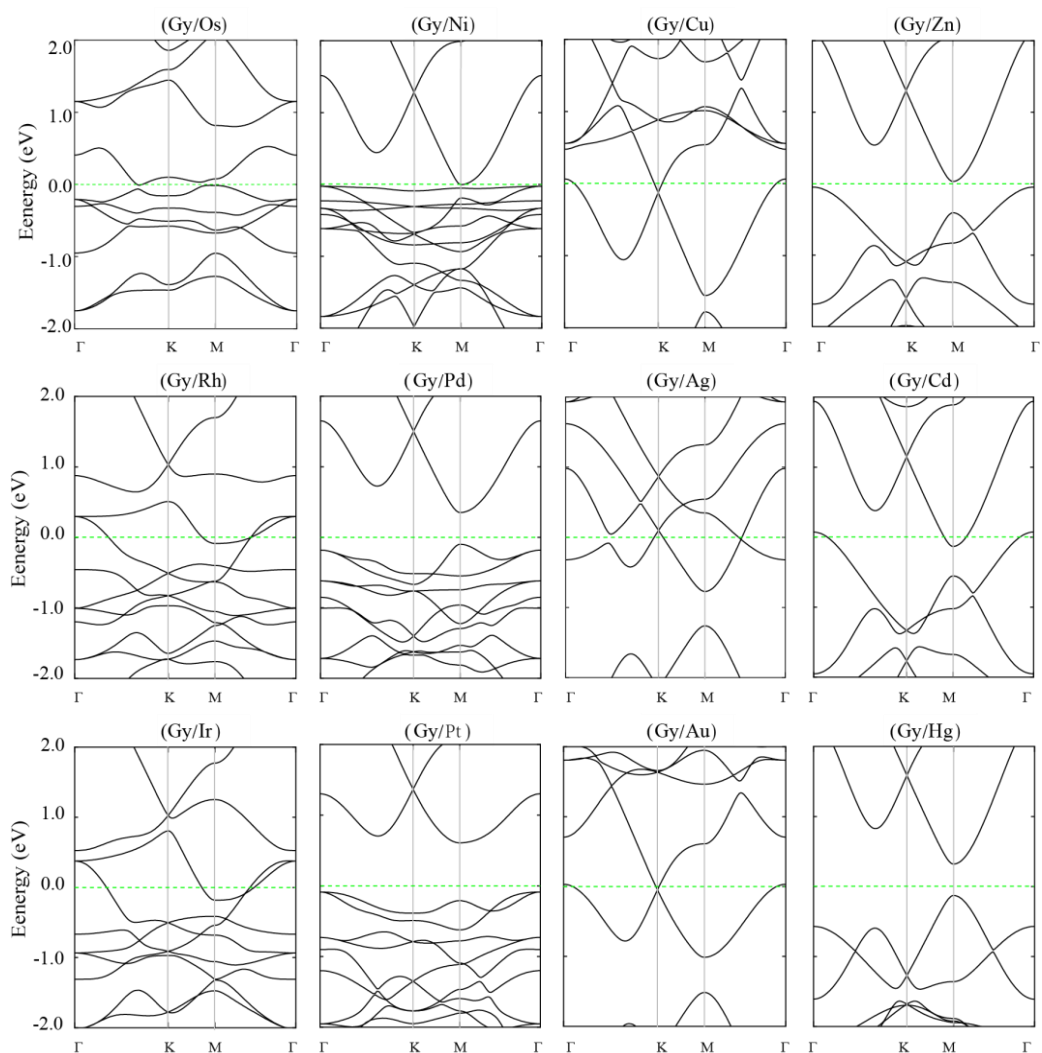

**Figure S3.3 The band structure of (a) the antiferromagnetic ground state with magnetic moment of a single magnetic atom is less than 0.5μB and (b-l) non-magnetic system.** It's shown that the systems of Gy/Cu, Gy/Ag and Gy/Au have the properties of semi-metals.

## S4 The curie temperature of Gy/Mo and Gy/W

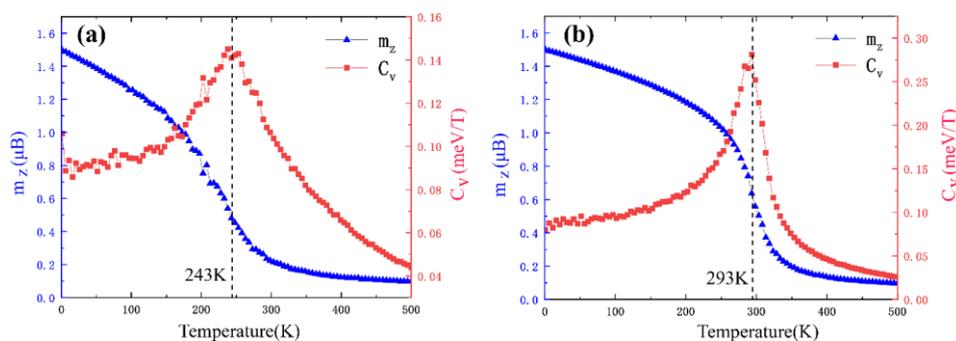

**Figure S4. The curie temperature of (a) Gy/Mo and (b) Gy/W.** Blue and red lines respectively represent the variation of the magnetic moment per magnetic atom of Gy/TM (TM=Mo, W), the specific heat capacity of the system with temperature were calculated by using the Monte Carlo method.

## S5 Spin charge density and super exchange interaction diagram of Gy/Cr, Gy/Mo and Gy/W.

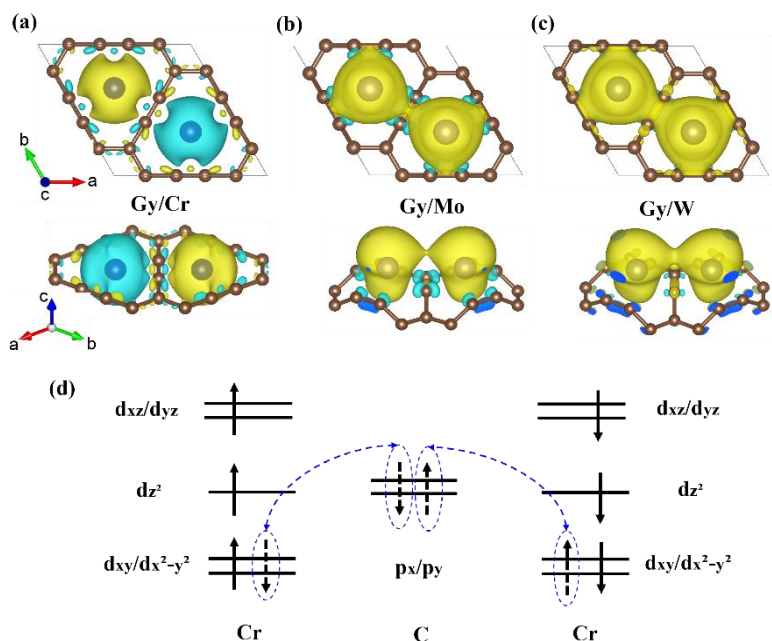

**Figure S5. Spin charge density (SCD) and super exchange interaction diagram of Gy/Cr, Gy/Mo and Gy/W.** (a), (b) and (c) are the SCD of Gy/Cr, Gy/Mo, and Gy/W, respectively. The spin-up and spin-down charge densities are represented by yellow and blue, respectively. (d) shows a schematic diagram for the super exchange interaction of the Gy/Cr. It can be seen that there are both super exchange and direct exchange interactions between TM atoms. The magnetism in Cy/Cr is mainly dominated by the super exchange interaction. While, the magnetisms in Gy/Mo and Gy/W are dominated by the direct exchange interaction, since the distances between TM atoms are smaller than that of Gy/Cr due to the deformation of Gy.

**S6 K-point test of Gy/Mo and Gy/W devices**

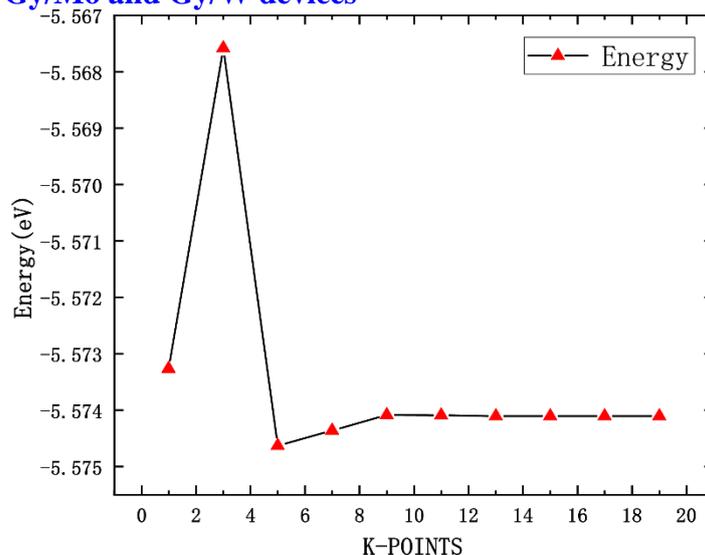

**Figure S6. K-point test of Gy/Mo and Gy/W devices.** We conducted a k-points test on the electrode of the device, and it can be seen from the figure that when K point is set to 15, the energy convergence requirement is met.

**S7 I-V curve of the Gy/Mo and Gy/W device.**

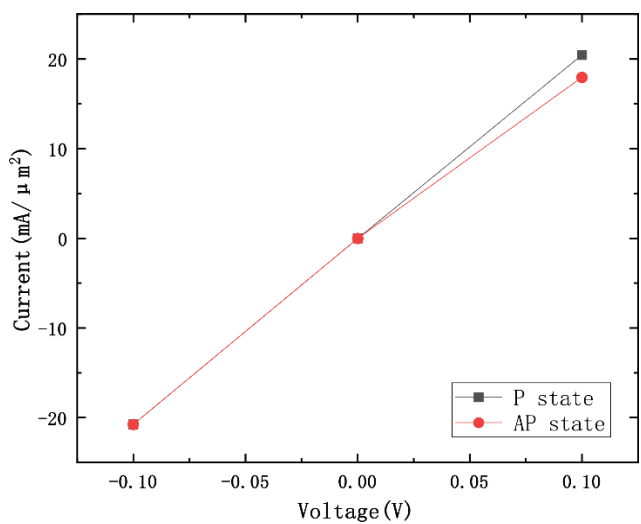

**Figure S7. I-V date of the device under A and AP state.** The red and black solid lines represent the I-V transfer curve in the parallel (P) state and the anti-parallel (AP) state, respectively. It is seen that the current density of them under 0.1 V are obviously arger than that in the vertical MTJ based of multilayer CrI3 [Nano Lett. 19, 915 (2019)]. This result indicates that the Gy/TM type MTJ is more superior in the low-power spintronic devices.